\documentclass[12pt]{iopart}
\usepackage{graphics} 
\begin{document}

\def\om0{{\bf\Omega}}
\def\g0{\bi{g}}
\def\spd#1{\left<#1\right>_f}

\title[Geophysical studies...]{Geophysical studies with
laser-beam detectors of gravitational waves}

\author{ L.Grishchuk\dag\ddag\ , V.Kulagin\ddag\ ,
V.Rudenko\ddag\ , A.Serdobolski\ddag}

\address{\dag\ School of Physics and Astronomy, Cardiff University, 
Cardiff CF24 3YB, United~Kingdom}

\address{\ddag\ Sternberg Astronomical Institute, Moscow State University,
Moscow, 119899, Russia}

\ead{grishchuk@astro.cf.ac.uk}

\begin{abstract}
The existing high technology laser-beam detectors of gravitational waves 
may find very useful applications in an unexpected area - geophysics. 
To make possible the detection of weak gravitational waves in the region of 
high frequencies of astrophysical interest, $\sim 30 - 10^3~ Hz$, control 
systems of laser interferometers must permanently monitor, record and 
compensate much larger external interventions that take place in the region
of low frequencies of geophysical interest, $\sim 10^{-5} - 3 \times~10^{-3}~
Hz$. Such phenomena as tidal perturbations of land and gravity, normal mode 
oscillations of Earth, oscillations of the inner core of Earth, etc. will 
inevitably affect the performance of the interferometers and, therefore, the 
information about them will be stored in the data of control systems. We 
specifically identify the low-frequency information contained in distances 
between the interferometer mirrors (deformation of Earth) and angles between 
the mirrors' suspensions (deviations of local gravity vectors and plumb lines). 
We show that the access to the angular information may require some modest 
amendments to the optical scheme of the interferometers, and we suggest the 
ways of doing that. The detailed evaluation of environmental and instrumental 
noises indicates that they will not prevent, even if only marginally, the 
detection of interesting geophysical phenomena. Gravitational-wave instruments 
seem to be capable of reaching, as a by-product of their continuous
operation, very ambitious geophysical goals, such as observation of the 
Earth's inner core oscillations.
\end{abstract}

\pacs{0480N, 0710F, 0150P, 0340K}

\submitto{\CQG}
 
\section{Introduction}

Laser-beam detectors of gravitational waves are designed to explore 
the Universe in a new type of radiation and from a 
new perspective. With the already operating instruments, 
and coming soon online, we are expecting to witness 
the discovery of fascinating physics involved in powerful sources of 
cosmic gravitational radiation. The astrophysical aims of the 
gravitational wave (g.w.) science are well understood and 
comprehensively described in the literature (for recent reviews, 
see for example \cite{thorne, gr5, cthorne, gr}). Naturally,
g.w. community is focused on the ambitious target of `reaching for the 
black holes'. But what about a more modest goal of looking inside our 
own planet~? No doubt, black holes are fascinating and scientifically 
important objects. But it is also necessary to remember that, say, the 
still poorly understood and unpredictable earthquakes are claiming 
thousands of human lives per year. Is it possible that the 
cutting-edge technology of g.w. interferometers \cite{ligoweb, GEOweb,
virgoweb, TAMAweb} may help us with accurate geophysical studies, 
as a by-product of the continuous search for astrophysical 
gravitational waves ? 

This is the major question of the present work, and the 
answer is positive. We shall show in this paper 
that a laser-beam detector of gravitational waves is in fact 
automatically a valuable geophysical device. Without collecting, 
recording and processing environmental information of geophysical
origin, a g.w. laser interferometer would simply be incapable of 
working as a sensitive astrophysical instrument. Certainly, we have to 
make sure that the extraction of geophysical information 
does not compromise the astrophysical aims of the instrument.
 
A laser-beam detector of gravitational waves is a conceptually simple
installation. Each arm of the interferometer, typically of the
length $L= 3~km$, consists of two mirrors, and the distance variation 
between the mirrors is monitored by the laser beam. The mirrors
are hanging on wires in supporting towers. Each mirror is  
essentially a mass element of the pendulum placed in the local 
gravitational field of the Earth. The eigen-frequency of the pendulum is
normally in the range of $0.1 - 1~Hz$. The multi-stage pendulum system 
shields the mirrors from large uncontrollable displacements of the 
tops of the supporting towers. This makes the interferometer capable of 
measuring, in the region of relatively high frequencies of 
astrophysical interest $\sim 30 - 10^3~ Hz$, the incredibly small 
variations of distance between the mirrors, at the level of 
$10^{-16}~ cm$. The expected cause of these variations is the 
incoming astrophysical gravitational wave. 

The isolation from noises in the region of relatively high
frequencies is only a part of the story. To ensure successful 
performance of the interferometer as an astrophysical instrument, 
control systems of the interferometer should also register and 
compensate for large external interventions of geophysical origin that 
take place in the region of relatively low frequencies. For example, the 
tidal half-daily variations of distance beween towers separated by $3~ km$ 
are typically at the level of $10^{-2}~ cm$. This change of distance is 
14 orders of magnitude larger than the anticipated astrophysical signal. 
If this sort of variations were allowed to affect distance between 
the mirrors, the interferometer would not be in the `locked' state,
and hence it would not be able to operate as astrophysical instrument. 
The `locking' of the interferometer requires that the distance between
the mirrors is maintained unchanged with accuracy of approximately one 
hundredth of the laser light wavelength, which amounts to 
$10^{-6}~ cm$ and less. This means that the low-frequency variations 
of distance between the mirrors should be monitored and largely 
removed by a control system called the adjustment system. 
A similar monitoring and compensation should be done with respect 
to low-frequency variations of the angle between the 
interferometer mirrors. This is being done by a control system 
called the alighnment system. Ideally, in order to reach the 
astrophysical goals, control systems should keep the interferometer 
in the working condition for the duration of time 
exceeding many months. 

Thus, the collected and recorded low-frequency information, 
which is vital for maintaining the operational state 
of the laser-beam detector of gravitational waves, 
inevitably makes the g.w. detector also a geophysical 
instrument. The time-scales of geophysical processes, some of 
which are believed to be crucial for global geodynamics, lie in the 
range from several minutes to several hours. In other words, we will be
interested in frequencies, which we call geophysical frequencies, 
somewhere in the interval $10^{-5} - 3 \times 10^{-3}~Hz$. 

In Sec.2 we consider a simple model of g.w. interferometer as
a geophysical instrument. It is assumed that the mirrors' suspension points 
can move and the plumb lines of hanging mirrors can vary. These changes
arise as a result of the Earth surface deformations and variations
of local gravitational field caused, for example, by the internal 
Earth dynamics. We derive general formulas for the distance between the
mirrors and the angles between the local plumb lines. Ideally, these are
two variables that are supposed to be monitored by, respectively, the 
adjustment and alignment systems of the interferometer.

In Sec.3 we consider a number of interesting geophysical phenomena which
inevitably affect the performance of a g.w. interferometer. The 
signatures of these phenomena are contained in the outcomes of the 
adjustment and alignment control systems. The geophysical effects to be 
studied include tidal perturbations, normal modes of Earth oscillations,
movements of the inner solid core of Earth, etc. We place the main
emphasis on the fascinating phenomenon of the inner core oscillations.
We estimate the useful geophysical signal accompanying this phenomenon, 
which will manifest itself in the variation of distance between the 
mirrors and in the variation of angle between the plumb lines of the
hanging mirrors. The guidance for the expected amplitude of the signal 
is provided by the reported in the literature indications that the 
inner core oscillations have been actually detected by other, traditional, 
methods. From the requirement that the signal to noise
ratio should be larger than 1, we define the level of tolerable 
noise in the proposed measurements. Specifically, the tolerable noise 
allows the detection of the useful signal, if the observation time 
exceeds 70, or so, inner core oscillation periods. 

A useful signal can be detected if the environmental and instrumental
noises are smaller than the calculated level of tolerable noise. In Sec.4 
we consider noises which we find most dangerous. We explicitely show 
that seismic, atmospheric and instrumental noises should not 
be capable of preventing the detection of inner core oscillations, 
even if only marginally. This refers both to distance and angle measurements.
There exists, however, a specific problem with the angle measurements,
related to the fact that the presently operating alignment systems 
are subject to a certain degeneracy. They cannot tell apart a tilt of the
mirror, which we are mostly interested in, and a latteral shift of the
mirror, which can be caused by a dull deformational noise. We analyze
this difficulty in great detail in Sec.5.

Since the angular measurements provide an important additional channel 
of geophysical information, we adress the problem of degeneracy in 
Sec.6 and suggest the ways of its circumvention. The desire to
keep the angular channel useful for geophysical applications may require 
some modest and harmless modifications of the optical scheme of g.w.
interferometers. We discuss at some length a few ideas with regard
to such modifications. It appears that, without interfering with
the astrophysical program of the instrument, certain geophysical
modifications are feasible.

In Sec.7 we emphasize some conclusions of the paper.

\section{Laser-beam detector of gravitational waves as a 
geophysical instrument}

To see better how a g.w. interferometer can work as a 
geophysical instrument, we will have to consider idealized, but 
representative, models. We start with a model of a pendulum, 
whose suspension point oscillates \cite{4a}, and which is placed in
a variable gravitational field. 

\begin{figure}
  \begin{center}
\includegraphics{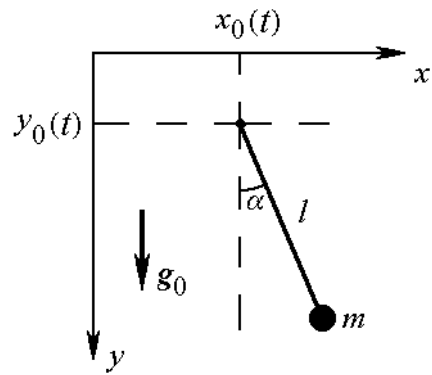}
  \end{center}
  \caption{\label{f0} A pendulum with a moving suspension point
  placed in a variable gravitational field.}
\end{figure}

A pendulum of mass $m$ and length $l$ can oscillate in the
$x, y$ plane, see Fig.\ref{f0}. Its suspension point has 
time-dependent coordinates $x_0(t)$, $y_0(t)$, and the local 
gravity acceleration vector ${\g0}(t)$ is also a function of 
time: ${\g0}(t)=[g_x(t),\ g_0+g_y(t)]$. The local gravity 
field is spatially homogeneous on the scale of small oscillations 
of the mass, but it is not homogeneous on larger scales. Later, we 
will take into account its inhomogeneity (spherical symmetry) on the 
scale of two widely separated pendula and even on the scale of 
displacements of the suspension point of an individual pendulum.

The only dynamical variable in this problem is the angle $\alpha(t)$ 
between the suspension wire and the axis $y$. The coordinates of 
the mass are given by

\begin{equation}
\label{eq0}
x(t)=x_0(t)+l\sin\alpha(t),\ \ y(t)=y_0(t)+l\cos\alpha(t).
\end{equation}

\noindent
The Lagrangian of the system can be written as
\begin{equation}
L=\frac{m}{2}(\dot x^2+\dot y^2)+mg_xx+m(g_0+g_y)y
\end{equation}

\noindent
Using expressions (\ref{eq0}) and ignoring non-dynamical
terms and total derivatives, one transforms the Lagrangian
to the final form: 

\begin{equation}
\label{finL} 
L=\frac{1}{2}ml^2\dot \alpha^2-ml(\ddot x_0\sin\alpha+\ddot
y_0\cos\alpha) +mg_x l\sin \alpha +m(g_0+g_y) l\cos \alpha.
\end{equation}

Now one can write down the equation of motion of the pendulum:  
\begin{equation}\label{dyneq0}
l\ddot\alpha=-\ddot x_0\cos\alpha+\ddot y_0\sin\alpha+g_x\cos\alpha
-(g_0+g_y)\sin\alpha.
\end{equation}
We assume that the time-dependent components of the gravity 
vector are very small, $g_x(t)\ll g_0$, $g_y(t)\ll g_0$, 
and the acceleration of the suspension point $\ddot y_0$ is much 
smaller than the unperturbed gravity acceleration $g_0$. 
We also assume that $\alpha\ll 1$, $\sin\alpha\simeq\alpha$,
$\cos\alpha\simeq 1$. Then, Eq. (\ref{dyneq0}) simplifies:
\begin{equation} \label{2}
\ddot\alpha+\omega_p^2\alpha=-\frac{\ddot x_0}l+\omega_p^2\alpha_0~,
\end{equation}
where $\omega^2_p=g_0/l$ is the square of oscillation frequency of 
the unperturbed pendulum. The angular variation of the 
local gravity vector $\g0$, that is, the angular deviation of the local
plumb line, is denoted $\alpha_0(t)=g_x(t)/g_0$. We see that, 
in the linear approximation in terms
of $\alpha$, the pendulum behaviour is affected only by changes in 
the gravity force direction, namely, by the component $g_x$ of the
gravity acceleration. The variations of the absolute 
value $|\g0|$ and the component $g_y$ of the acceleration 
produce smaller effects, which we neglect.

Equation (\ref{2}) is an inhomogeneous linear differential equation 
with constant coefficients. In the frequency domain, the relevant solution
can be written
\begin{equation}
\alpha_{\omega}= \frac{\omega^2\,x_{0\omega} + l\omega_p^2
\alpha_{0\omega}}{l\,(\omega_p^2-\omega^2)}, 
\label{3}
\end{equation}
where $\alpha_{\omega},\, x_{0\omega},\,\alpha_{0\omega}$ are 
complex Fourier amplitudes of the corresponding variables at 
frequency $\omega$. The divergent amplitude at the resonance
frequency $\omega = \omega_p$ will be tempered by friction, which
we have ignored. The regions of our interest are 
relatively high frequencies $\omega \gg\omega_p$ for
astrophysical applications and relatively low frequencies 
$\omega \ll \omega_p$ for geophysical applications. For orientation,
one can think of $\omega_p/2 \pi = f_p \approx (0.1 - 1)~ Hz$. This is 
true for the existing instruments.  

{\it a) High frequencies $\omega \gg
\omega_p$ -- astrophysical applications}. \\
The pendulum angular amplitude is given by 
\begin{equation} \label{4a}
\alpha_{\omega}\simeq \left[\left(\frac{\omega_p}{\omega}
\right)^2-1 \right]^{-1} \frac{x_{0\omega}}{l}- \left(\frac{\omega_p}
{\omega}\right)^2 \alpha_{0\omega} \simeq
- \frac{x_{0\omega}}{l}- \left(\frac{\omega_p}
{\omega}\right)^2 \alpha_{0\omega} ~.  
\end{equation}
The horizontal position of the pendulum mass is given by  
\begin{eqnarray} 
\label{5}
x_{\omega} & \simeq & x_{0\omega}+ l\alpha_{\omega} \simeq 
x_{0\omega} \left[1 + \frac{1}{(\omega_p/\omega)^2 - 1}\right] -
l \alpha_{0\omega}(\omega_p/\omega)^2 \nonumber \\ 
& \simeq & -\left(\frac{\omega_p}{\omega}\right)^2 
(x_{0\omega}+ l\alpha_{0\omega}) 
 =  -\left(\frac{\omega_p}{\omega}\right)^2 x_{0\omega}- 
\frac{g_{x\omega}}{\omega^2}. 
\end{eqnarray}
Eq.(\ref{5}) illustrates the well known method of vibrational 
isolation of interferometer's test masses-mirrors. The residual 
displacement $x_{\omega}$ of the pendulum mass is a factor 
$(\omega_{p}/\omega)^2$ smaller than the displacement 
$x_{0\omega}$ of the suspension point. The $N$ stages of isolation
reduce the external $x_{0\omega}$ by a factor $(\omega_{p}/\omega)^{2N}$.
This makes the mirror shielded from large deformational shifts 
of the suspension point, and therefore the interferometer
becomes capable of detecting increadibly small distance variations 
caused by astrophysical signals. 
However, the displacement caused by the local gravitational
environment (the last term in Eq.(\ref{5})), cannot be shielded.
If this gravitational noise is present, it creates certain 
problems for the astrophysical program, which we will return to later.

{\it b) Low frequencies $\omega \ll \omega_p$
-- geophysical applications }. \\
The pendulum angular amplitude is given by 
\begin{equation} 
\label{4}
\alpha_{\omega}\simeq \left(\frac{\omega}{\omega_p}\right)^2
\frac{x_{0\omega}}{l}+ \alpha_{0\omega}. 
\end{equation}
The contribution to $\alpha_{\omega}$ provided by movements 
of the suspension point (first term in Eq.(\ref{4})) is
suppressed by a factor $(\omega/\omega_p)^2$. At sufficiently
low frequencies, this term can become smaller than  
the second term in Eq.(\ref{4}). One can say that,
in this regime, the equilibrium position of the pendulum follows 
a new plumb line $\alpha_{0\omega}$ determined by the 
slow variation $g_x(t)$ of the local gravity force vector $\g0$.

The horizontal position of the pendulum mass, in the low-frequency
approximation, is given by 
\begin{eqnarray} \label{5new}
x_{\omega}\simeq x_{0\omega}+ l\alpha_{\omega} \simeq 
x_{0\omega} +\left(\frac{\omega}{\omega_p}\right)^2 x_{0\omega}+ 
l\alpha_{0\omega} \simeq x_{0\omega} + l\alpha_{0\omega}~.
\end{eqnarray}
It follows from Eq.(\ref{5new}) that the movements of the suspension
point transmit practically without change into the movements of the 
pendulum mass. In other words, the low-frequency deformations cannot
be shielded, so they should be monitored and compensated by the 
control systems.  

\begin{figure}
  \begin{center}
\includegraphics{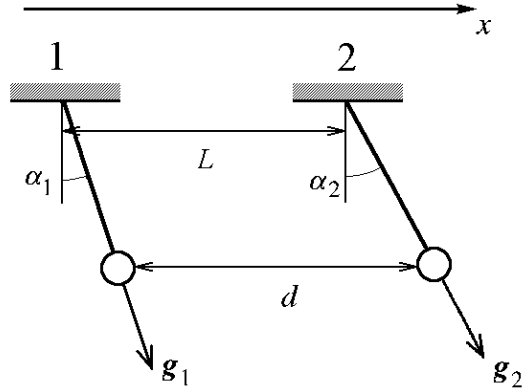}
  \end{center}
  \caption{\label{f1} Relative variations of plumb lines and mass`
positions in an interferometer.}
\end{figure}

The interferometer arm includes two separated pendula, and 
it is the difference of displacements and difference of mirrors'
inclinations that are actually being monitored. We assume that the 
pendula are identical, but their environments are different,
see Fig.\ref{f1}. 
We denote by symbol $\delta$ the variations at each site
and by symbol $\Delta$ the difference of variations at two sites.
It is important to know how different conditions are at two sites.
If the characteristic spatial scale $\lambda$ of perturbations is 
shorter than $L$, then the conditions at two sites are not correlated,
and $\Delta$-variations are typically of the same order of magnitude 
as the largest of individual $\delta$-variations. However, if both sites 
are covered by a long-wavelength perturbation $\lambda \gg L$, then
the $\Delta$-variations are much smaller than the $\delta$-variations.
Symbolically, we can write 
\begin{equation}
\Delta \simeq \frac{L}{\lambda} \delta,~~~~~~~~ {\rm if}
~~~~~~~~ L \ll \lambda. 
\label{8} 
\end{equation}

Let the unperturbed coordinates of the suspension points be 
$x_{01},\,x_{02}$. Each 
point is subject to time-dependent deformational shifts 
$\delta{x_{01}},\,\delta{x_{02}}$. Of course, they are
much smaller than the unperturbed distance $L$, $L =x_{02}-x_{01}$.
We will also use the difference of the deformational shifts, 
$\Delta{x_0}(t)= \Delta L = \delta{x_{02}}(t)-\delta{x_{01}}(t)$.
The difference of angular variables of the two pendula is
$\Delta \alpha(t) = \alpha_2(t) -\alpha_1(t)$. The difference of local
values of $g_x$ and, hence, the angle between the local plumb lines 
at two sites is $\Delta \alpha_0(t)=\alpha_{02}(t) -\alpha_{01}(t)$. 
[Here we ignore the fact that the Earth's gravitational field is
centrally-symmetric, which leads to slightly different orientations
of the unperturbed gravity force vectors at two cites, but later 
we will take this fact into account.] Writing
equations (\ref{2}) at two sites, and taking their difference,
we arrive at the equation 
\begin{equation}
\ddot{\Delta \alpha}+\omega_p^{2} \Delta \alpha = 
-\frac{\ddot{\Delta x_0}}{l}+ \omega_{p}^{2}\Delta \alpha_0~.  
\label{6}
\end{equation}
Certainly, the relevant solution to this equation is given by
the formula similar to (\ref{3}), with obvious substitutions: 
\begin{equation}
\label{6a}
\Delta \alpha_{\omega}= \frac{\omega^2\,\Delta x_{0\omega} + l\omega_p^2
\Delta \alpha_{0\omega}}{l\,(\omega_p^2-\omega^2)}. 
\end{equation}
The variation of distance between the masses is 
\begin{equation}
\Delta d_{\omega} = \Delta{x_{0 \omega}} + l \Delta \alpha_{\omega}. 
\label{7}
\end{equation}

The high-frequency and low-frequency approximations to equations
(\ref{6a}), (\ref{7}) follow the lines of equations (\ref{4a}), 
(\ref{5}), (\ref{4}), (\ref{5new}). In particular, in the low-frequency
region, $\omega \ll \omega_p$, we have
\begin{equation} \label{9}
\Delta \alpha_{\omega}\simeq \Delta \alpha_{0\omega}+ 
\left(\frac{\omega}{\omega_p}\right)^2
\frac{\Delta x_{0\omega}}{l} 
\end{equation}
and
\begin{eqnarray} \label{10}
\Delta d_{\omega} \simeq \Delta x_{0\omega} + l \Delta \alpha_{0\omega}~.
\end{eqnarray}
It is seen from Eq.(\ref{10}) that the large deformational
contribution $\Delta x_{0\omega}$ leaks into the system. 
This is why the interferometer's adjustment system records and 
largely removes $\Delta x_{0\omega}$ from $\Delta d_{\omega}$ by 
applying force to the mirrors and mechanically adjusting their 
positions. (The adjustment system also tunes the frequency of the 
laser light). The circuits of adjustment system is one of the 
sources of geophysical information. 

The angle $\Delta \alpha_{\omega}$
between the suspension wires, Eq.(\ref{9}), is mostly the
reflection of the angle between local plumb lines, first term
in Eq.(\ref{9}). For sufficiently low frequencies, the second
term in Eq.(\ref{9}) is smaller than the first one. 
We assume that the angle between the wires is at the same time 
the angle between surfaces of the mirrors. Under the condition
that other contributions to Eq.(\ref{9}) can be 
identified and separated (one complication, related
to the centrally-symmetric character of the Earth's gravitational
field, is treated in detail in Sec.4c), the alighnment 
system, whose purpose is to monitor the angle between the mirrors, 
will be monitoring the angle $\Delta \alpha_{0\omega}$ between 
the plumb lines. This control system is another source of 
geophysical information. 

Variables of geophysical interest, $\Delta x_{0\omega}$ and
$\Delta \alpha_{0\omega}$, consist of two parts. 
One part is a useful geophysical signal, another 
part is environmental noise. In general, temporarily ignoring 
noises and other complications, the knowledge of the output 
signals $\Delta \alpha_{\omega}$ and $\Delta d_{\omega}$ allows 
one to solve equations (\ref{9}), (\ref{10}) with respect to 
$\Delta x_{0\omega}$ and $\Delta \alpha_{0\omega}$:
\begin{equation} \label{11}
\Delta x_{0\omega} \simeq \Delta d_{\omega}- l \Delta \alpha_{\omega}~,~~~~~~ 
\Delta \alpha_{0\omega} \simeq \Delta \alpha_{\omega}- 
\left(\frac{\omega}{\omega_p}\right)^2 \frac{\Delta d_{\omega}}{l}~.
\end{equation}
Ideally, this would be a reconstruction, as complete as possible,
of the acting geophysical perturbation. However, 
complications do exist, and we will perform more detailed 
analysis in the rest of the paper.   

In the end of this section we shall briefly discuss the
distance variations and plumb line deflections in terms of the 
Newtonian gravitational potential $U({\bf x},t)$ of the Earth.
The potential $U$ is the sum of the unperturbed potential
$U_0 = GM/r$ and perturbations caused by external and internal
gravitational fields. The external fields are mostly the 
tidal effects of the Moon, Sun and planets. The internal fields 
include such effects as normal oscillation modes of Earth and 
oscillations of the 
inner core of Earth. At the surface of an idealized rigid body, 
the local gravitational acceleration would be given by
${\bf g} = \partial U/ \partial {\bf x}$. The true value of
${\bf g}$ at the surface of real elastic Earth is somewhat 
different from this quantity. This happens because the 
gravitational field perturbations, originating from either external 
or internal sources, displace elements of the Earth's surface
in its own gravitational field and also give rise to a 
redistribution of the Earth's matter density.
In the point of observations, these changes create additional 
variations of the gravitational potential. In conventional geophysics 
these effects are taken into account by some numerical factors 
called the Love numbers \cite{c12}. For a perfectly rigid
body the Love numbers are zeros, while for a fluid body they are 
ones. On real deformable Earth, the Love numbers are roughly 
of the order of 1.  

The difference of variations of ${\bf g}$ at two sites separated
by $L$ is causing a change of distance between the sites, and
it also determines a relative deflection of local plumb lines. 
Let the common unperturbed ${\bf g}_0$ point out in the $y$ direction,
and the sites be located on the $x$ axis. Then, if $L$ is small
in comparison with the characteristic length $\lambda$ at which the
perturbation of the gravitational potential changes, one can derive: 
\begin{equation}
\Delta \alpha_0 = \frac{L}{g_0} 
\sqrt{\left( \frac{\partial g_x}{\partial x} \right)^2 +  
\left( \frac{\partial g_z}{\partial x} \right)^2 }=
\frac{L}{g_0}   
\sqrt{\left( \frac{\partial^2 U}{\partial x^2} \right)^2 +  
\left( \frac{\partial^2 U}{\partial z \partial x} \right)^2}.  
\label{pot}
\end{equation}
It is clear that we are dealing with gradients of gravity 
variations $\delta {\bf g}$. A projection of this formula on 
the $x,y$ plane returns us to 
the results discussed above (compare with Eq.(\ref{8})):
\begin{equation}
\Delta \alpha_0 = \frac{L}{g_0} \frac{\partial g_x}{\partial x}
\approx \frac{\delta g}{g} \frac{L}{\lambda} \approx \delta\alpha
\frac{L}{\lambda}. 
\label{8a}
\end{equation}

\section{Geophysical phenomena to be studied}
\label{s4}

A number of interesting geophysical phenomena are in fact 
geophysical signals that can be studied with the
help of laser-beam detectors of gravitational waves. The division
of geophysical phenomena into signals and noises is not 
clear-cut. For example, the low-frequency seismic perturbations 
can be assigned to either category. Likewise, periodic thermal 
deformations of land are probably the largest ones numerically, but 
they are not of gravitational origin, and we treat them as a 
predictable and removable hindrance rather than an interesting 
geophysical signal. In general, we shall regard signals some 
quasi-periodic processes with strong participation of gravity in them, 
whereas more random processes we regard noises, hampering detection of 
the signals. Noises are considered in the next section, while here we
focus on some quasi-periodic phenomena which are believed to be 
important for global geodynamics. They are accompanied 
by variations of local gravity vector and land deformations. Both 
of these perturbations influence the performance of laser 
interferometric detectors of gravitational waves, and should be 
monitored and removed by the control systems.

\medskip
\noindent {\it a). The Earth tides.}
\medskip

Large perturbations of gravity ${\bf g}$ and deformations of land 
${\bf u}$ are induced by the gravitational fields of the Moon and, 
to a smaller extent, Sun. The main tidal harmonics are well 
described and measured \cite{c12}. The values and directions of the 
perturbed quantities depend on the position of the observation point 
on the Earth surface, but we will be mostly interested in their typical 
(not exceptionally small and not exceptionally large) numerical  
amplitudes. The dominant contributor to the tidal effects is the 
lunar $M_2$ harmonic which has a period of $\tau\simeq 12~ hours$ 
and which is quadrupolar in terms of its angular dependence. 
The typical $M_2$ amplitude of the variations of the absolute
value of ${\bf g}$ is $\delta g\simeq 45~\mu Gal$ 
($ Gal =  cm/sec^2,~\mu Gal = 10^{-6}~Gal,~nGal = 10^{-9}~Gal$). 
The unperturbed gravity field is
$g \simeq 980~ Gal$, so that for the $M_2$ harmonic we 
have $\delta g/g \simeq 5 \times 10^{-8}$. 
The main solar harmonic is $S_2$ with $\tau = 12~ hours$ and 
$\delta g \simeq 20~\mu Gal$. The higher frequency tidal 
harmonics are considerably weaker. For example:
$M_{3}$ with $\tau \simeq 8~ hours$, $\delta g \simeq 0.8 ~\mu Gal$ and 
$M_4$ with $\tau \simeq 6 ~hours$, $\delta g = 0.03 ~\mu Gal$. 

On the Earth surface, variations of gravity ${\bf g}$ and 
displacements ${\bf u}$ of land elements 
are linked by the Love numbers. If $W_2$ is 
the quadrupole component of the tidal potential, one can write
for the radial components $\delta g_r$ and $u_r$ \cite{c12}: 
\begin{equation}
\label{tides1}
\delta g_r = 2(1+h-\frac{3}{2} k) \frac{W_2}{R_e},
\end{equation} 
\begin{equation}
\label{tides2}
u_r =\frac{h}{g} W_2,
\end{equation}  
where $R_e$ is the Earth radius ($R_e \approx 6400~ km$) and $h, k$ are
the Love numbers ($h \approx 0.6,~ k \approx 0.3$). Since the radial and 
angular components of the displacement vector ${\bf u}$ are approximately
equal, we will be using for all of them the approximate relationship
\begin{equation}
\label{tides3}
u \approx 0.3~ \frac{\delta g}{g} R_e.
\end{equation}
This relationship is valid for any gravitational perturbation, not
necessarily a tidal one. Specifically for the $M_2$ tidal component,
a typical displacement amounts to $u \approx 10~ cm$. 
The observed maximal displacements are a factor $3 - 5$ larger than 
this evaluation. 

The characteristic spatial scale of the considered quadrupole 
perturbation is $\lambda \simeq R_e$. Therefore, assuming that 
the armlength of the interferometer is $L = 3~ km$, the 
$\Delta$-variation is smaller than the $\delta$-variation 
(see Eq.(\ref{8})) in proportion to $L/\lambda$, which reduces to 
\begin{equation}
\label{difffactor}
\frac{L}{R_e} \approx 5 \times 10^{-4}. 
\end{equation}
Combining the numbers, we arrive at  
\begin{eqnarray}
\label{tides4}
\Delta x_0 & = & u \frac{L}{R_e} =0.3~\frac{\delta g}{g} L 
\approx 5 \times 10^{-3}~ cm,~~~~~~{\rm and} \nonumber\\ 
\Delta \alpha_0 & = & \frac{\delta g}{g} \frac{L}{R_e} \approx 
3 \times 10^{-11}~ {\rm rad}.
\end{eqnarray}

As was already mentioned, tidal perturbations of this level of
magnitude should be easily seen by the interferometer 
control systems. A successful performance of the 
interferometer requires a much better 
precision in monitoring the distance between mirrors, at the
level not less than $(10^{-6} - 10^{-7})~ cm$. The danger of 
tidal effects for the LIGO interferometers and the 
necessity of their removal have been long recognised \cite{raab}. 
We focus on the positive side of this necessary procedure: 
the existing network of interferometers, located
in different parts of the globe, will allow one to obtain useful 
information on amplitudes, spatial patterns and temporal phases of 
tidal perturbations. This collected and analyzed low-frequency 
information will help refine numerical values of 
Earth's parameters, such, for example, as its Love numbers. 
 
\medskip
\noindent {\it b). Free oscillations of Earth.}
\medskip

As every elastic body, our planet is capable of free oscillations
(see, for example, \cite{bullen}). In the
approximation of a non-rotating, uniform, spherically-symmetric Earth,
it is convenient to expand perturbations in terms of 
spherical harmonics $Y_{lm}(\theta, \phi)$, where $\phi$ is longitude
and $\theta$ is co-latitude:    
\begin{equation}
u(r,\theta,\phi) \sim~~  {_{n}R_{l} (kr)}\, Y_{lm}(\theta, \phi) \,
\exp\left( i \omega t\right). 
\end{equation}
The discrete values of the wave number $k$, $k_{ln} = 
\pi \alpha_{ln} / R_e$, are determined by the radial boundary 
conditions and the $n$-th root, $\alpha_{ln}$, of the 
corresponding boundary equation (see, for example, \cite{mf}). 
The eigen-frequencies $\omega$ do not depend on $m$ and 
are given by $\omega_{ln} = \pi \alpha_{ln} c_s /R_e$, where $c_s$ 
is the appropriate speed of sound. If the multipole order $l$ is much 
smaller than a (nonzero) $n$, the deformations have maximal amplitudes 
in the central region of the sphere. In the opposite case of $l$ being
much larger than $n$, the deformations are mostly concentrated near 
the surface of the sphere, and it also holds that 
$\pi \alpha _{ln} \approx l$. In the extreme regime $l \gg n$, 
the deformations look more like seismic perturbations rather than free
oscillations of Earth. The short-wavelength seismic perturbations
can propagate deep to the central regions of Earth, refract and 
reflect back. This is an important diagnostic of the central regions.     
  
There are two basic types of elastic deformations. One is accompanied 
by variations of volume, matter density and gravity ${\bf g}$,
and is called spheroidal modes $_{n}S_{lm}$. Another
is not accompanied by these variations; it represents purely twisting
motions, and is called toroidal (shear) modes $_{n}T_{lm}$.
Some of the shear perturbations cannot propagate in liquids, and their
observed attenuation in the central regions of Earth helped discover the
liquid (outer) core. Many hundreds of Earth's normal modes 
are now identified from the study of ground
motions and gravimeter records. The longest period normal mode is
$_0S_2$ with period $T=53.8~ min$. This period is of course 
consistent with our evaluations given above. Taking 
$c_s \approx 4~ km/sec$ and $\alpha_{20} \approx 1$, we arrive
at the right value for $T$. The frequency of the $_0S_2$ 
mode is $f \approx 3\times 10^{-4}~ Hz$, whereas the frequencies of 
higher order $S$ and $T$ modes extend to the $milli~Hz$ region. For 
example, the $_0S_{20} - {_0S_{40}}$ modes ($l = 20 - 40$) have equally 
spaced frequencies in the interval of approximately
$(1 - 5)~ mHz$, again in agreement with the evaluations given above.    

Free oscillations of Earth are best observed after large earthquakes.
In contrast to tidal harmonics, the amplitudes of free Earth
oscillations cannot be predicted in advance. However, it is
known that typical observed values of gravity variations in the 
fundamental $_0S_2$ mode are at the level of about 1 
$\mu Gal$ \cite{15a, 15b, 16}. The amplitudes of higher 
frequency registered modes are smaller, but they can reach 
$\delta g\sim 0.1$ $\mu Gal$. The amplitudes of modestly
excited modes between $_0S_{20}$ and $_0S_{40}$ were detected in the
long series of observations on seismically quiet days. The
acceleration amplitudes are at the level of $2~ nGal$, and 
higher \cite{tu}. There exists the permanently present noise 
in these modes with the amplitudes corresponding to $0.4~ nGal$. 
Interestingly, this background level of land deformations 
and gravity variations cannot be explained by cumulative effect 
of small earthquakes, despite the fact that there occur 
thousands of them per year. It is suggested \cite{tu} to seek 
the source of these continuous oscillations of Earth in the action 
of atmospheric pressure variations on the Earth surface. We will
return to this discussion in our analysis of atmospheric
noises in Sec.4b. 

The modestly excited higher frequency $S$ and $T$ modes seem to be 
within the reach of adjustment and alighnment systems of the
gravity-wave detectors. The $T$ modes do not affect the plumb 
line directions, but they are present in the distance variations.  
Let us take a representative normal mode with 
$f = 4 \times 10^{-3}~ Hz$ and the amplitude of gravity 
variations $\delta g = 2~ nGal$, i.e. $\delta g/g = 2 \times 10^{-12}$.
The wavelength $\lambda$ is $\lambda = c_s/ f \approx 10^{8}~ cm$,
that is, the spatial scale of perturbations $\lambda$ is smaller
than $R_e$, but still much longer than the armlength $L=3~ km$ of the 
interferometer. 

We shall first generalise Eqs. (\ref{tides3}), (\ref{tides4})
to the case, where the spatial scale of $\delta g$ is $\lambda$, not
$R_e$. Then, we will have to write 
\begin{equation} 
\label{gen}
\Delta x_0  \approx  0.3~\frac{\delta g}{g} R_e \frac{L}{\lambda}~~~~~~
{\rm and}~~~~~~ 
\Delta \alpha_0  \approx  \frac{\delta g}{g} \frac{L}{\lambda}. 
\end{equation}
Note that, on the surface of deformable Earth, there exists a universal 
relationship between the distance and angle variations, which directly
follows from Eq.(\ref{gen}):
\begin{equation}
\label{coupl}
\Delta \alpha_0 \approx \frac{\Delta x_0} {0.3~ R_e}. 
\end{equation} 
Substituting the numbers in Eq.(\ref{gen}), we obtain the
following evaluations for the considered normal mode of oscillations
at $f = 4 \times 10^{-3}~Hz$:
\begin{equation}
\label{oscil}
\Delta x_0 \approx 10^{-6}~ cm,~~~~~{\rm and}~~~~~~
\Delta \alpha_0 \approx 6 \times 10^{-15}~ rad.
\end{equation}

Obviously, if the amplitude of the excited mode is larger 
than $2~ nGal$, this will increase the expected 
amplitudes of $\Delta x_0$ and $\Delta \alpha_0$.
In addition, normal modes, as any other oscillations, are characterised
by certain quality factor $Q$. The mode can execute $Q$ cycles 
before its amplitude degrades by a factor of 3, or so. 
If the observation of the mode lasts for at least as long as 
its relaxation time $\tau$, $\tau = Q/ \omega$, the signal to noise 
ratio increases in proportion to $\sqrt{Q}$. This makes the 
prospects of detection of the Earth normal modes much better. 
The quality factor of individual modes is uncertain, but the 
literature often quotes the numbers 
well in excess of 100 (see, for example, \cite{c11}).     

\medskip
\noindent {\it c). The Earth inner core oscillations}
\medskip

The oscillations of the real Earth are complicated by Earth's aspherical
shape, its rotation, the presence of gravitational restoring forces
in addition to elastic forces, the stratification and variation 
of material properties with depth, etc. Very important modifications
are caused by the existence of the Earth core (see, for example,
\cite{c11}). It is believed that the inner core is solid and has
radius $r_s \approx 1220~ km$ and density $\rho_s \approx
13~ g/cm^3$. The outer core is liquid and has radius 
$r_l \approx 3470~ km$ and density $\rho_l \approx 12.6~ g/cm^3$. 
[The current geophysical models claim a much higher 
precision of the Earth parameters than is actually needed 
for our evaluations.] 

The solid core is held in its equilibrium position within the
liquid core mainly by gravitational forces. It is reasonable
to expect that an earthquake, or some other cause, can 
occasionally excite oscillations of the inner core, predominantly 
along the Earth rotation axis (the polar mode) \cite{16, 17}. 
A displacement of a solid sphere within a liquid sphere is 
equivalent to a displacement of the effective mass 
$m = (4 \pi/3) (\rho_s -\rho_l) r_s^{~3}$. If the amplitude of 
the displacement is $A$, then the associated variation of 
gravity at the Earth surface is
\[
\delta g = \frac{2Gm}{R_e^3} A = \frac{8\pi}{3} G A (\rho_s -\rho_l)
\left (\frac{r_s}{R_e}\right)^3.
\]      
The evaluation of this expression using the parameters mentioned 
above gives
\[
\delta g \approx 2~ nGal~ \left(\frac{A}{cm}\right).
\]

Period $T_p$ of polar oscillations of the inner core mass 
$m_s$ is mainly determined by the gravitational restoring force:
\[
m_s {\ddot A} = - G \frac{m~\frac{4}{3} \pi \rho_l A^3}{A^2},
\]
which leads to $T_p^2 = 3 \pi \rho_s / G \rho_l (\rho_s - \rho_l)$.  
Corrections to this expression are taken into account by a
small numerical factor $\alpha_p$ \cite{17}, so that 
\begin{equation}
\label{ff1}
T_p \simeq \sqrt{ \frac{3\pi(\rho_s +\alpha_p \rho_l)} 
{G \rho_l(\rho_s -\rho_l)}}.   
\end{equation}
The evaluation of $T_p$ leads to a period of about 4 hours.
The Coriolis force of the rotating Earth splits this period
inito a triplet of periods, with the size of splitting 
determined by the Earth angular velocity.

The fascinating phenomenon of the inner core oscillations
will be in the center of our further discussion.
There exist indications that the inner core translational 
triplet has been actually detected \cite{18, 5aa}.
The analysis of long records from superconducting
gravimeters has shown three resonances with the central period
$T_C \approx 3.8~ hours$ and the side periods separated
from the central one by $ \sim \pm 10~ min$. The
amplitudes of each of the three oscillations are at the
level of $(5 - 6) ~nGal$. The quality factor $Q$ of each
resonance is somewhat higher than 100.    

Using evaluations similar to those that have been done in
previous subsections, we can estimate the effect of the
Earth inner core oscillations on the laser-beam detectors of
gravitational waves. We take the amplitude of gravity 
variations $\delta g/g = 6 \times 10^{-12}$,
the charecteristic spatial scale of gravity variations and land
deformations $\lambda_c \simeq 2~ R_e$, and the length of the 
interferometer arm $L = 3~ km$. Then, the distance and angular 
signal amplitudes amount to
\begin{equation}
\label{core}
\Delta x_0 \approx 3 \times 10^{-7}~ cm,~~~~~{\rm and}~~~~~~
\Delta \alpha_0 \approx 2 \times 10^{-15}~ rad.
\end{equation}

The detectability condition requires that the signal to noise
ratio $S/N$ should be better than 1. The resonance bandwidth of 
the core oscillations is $\Delta f = f_C /Q$, where the 
central resonance frequency is $f_C = 7.3 \times 10^{-5}~ Hz$.
We assume that the observation time of the signal 
is at least as long as $Q$ cycles of the signal. 
This means that the tolerable 
spectral noise ${\tilde N}$ per $\sqrt{Hz}$ is related to 
the signal amplitude $S$ according to
\begin{equation}
\label{sn}  
{\tilde N} ~ \sqrt{\Delta f} = N < S.
\end{equation}
Although the quality factor $Q$ for the Earth inner core 
oscilllations is somewhat larger than 100, we take it as $Q= 73$ 
in order to operate with the round number: $\sqrt{f_C/Q} = 10^{-3}$. 

Thus, we derive the allowed broadband 
noises in the region of resonances:
\begin{eqnarray}
\label{noises}  
{\Delta x}_{noise} \approx 3 \times 10^{-4}~ \frac{cm}{\sqrt{Hz}}, & & ~~~~~~ 
{\Delta \alpha}_{noise} \approx 2 \times 10^{-12}~ 
\frac{rad}{\sqrt{Hz}}, \nonumber\\  
& & {\delta g}_{noise} \approx 6 ~\frac{\mu Gal}{\sqrt{Hz}}. 
\end{eqnarray}
The relationship between the noise variables in Eq.(\ref{noises}) 
is of course the same as the relationship between the signal 
variables, Eq.(\ref{gen}) with $\lambda \simeq R_e$.  

Obviously, if for some reason the inner core gets excited to the
level much higher than $5 - 6~  nGal$, the requirements on the tolerable
noise could be significantly relaxed. Alternatively, if the noise 
remains at the level of Eq.(\ref{noises}), the $S/N$ gets much 
higher than 1.
For example, there exists evidence \cite{16} that the amplitude
of the inner core oscillations can be as large as $0.64~ \mu Gal$,
which would make the signal easily detectable. And, in any case, 
one can further improve $S/N$ by observing for longer time than 
the assumed 73 cycles.

\begin{figure}
  \begin{center}
\includegraphics{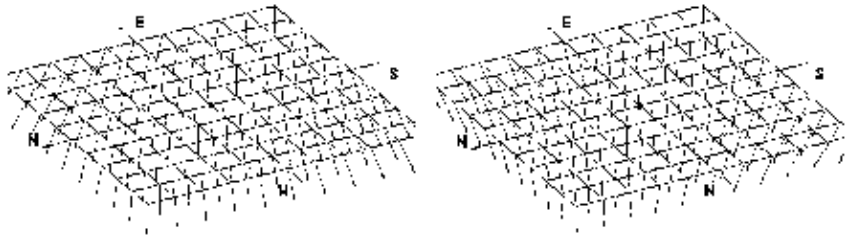}
  \end{center}
  \caption{\label{vectors} Gravity vectors describing tides and 
oscillations of the Earth inner core. Two distributions shown
on the panels are separated by a half of the tidal period.}
\end{figure}

Variations of gravity on the Earth surface within a patch of 
the size $L=3~km$ can be numerically simulated \cite{1a}. 
Fig.\ref{vectors} is the illustration of the gravity field
vectors disturbed by the inner core polar motion and 
by semidiurnal tides. To make the core contribution 
distinguishable on the graph, it was artificially 
enhanced by a factor 1000.

\section{Environmental and instrumental noises}

The important geophysical phenomena listed in the previous 
section will certainly affect the laser-beam detectors of 
gravitational waves. However, they can only be measured if 
the signals are not swamped by various noises. In principle, 
any random fluctuations of the Earth surface and any random 
redistributions of the surrounding masses of soil, air and water
can alter the distances and plumb line directions, which we are 
interested in. In addition to these environmental noises, there 
always exist imperfections in the measuring devices themselves, 
that is, instrumental noises. Here, we will analyse the noises 
which we consider most dangerous for the proposed measurements.

\medskip
\noindent {\it a) Seismic perturbations}
\medskip

It has been long recognized that seismic waves propagating in the 
vicinity of a laser interferometer can limit its performance 
as a gravitational wave detector \cite{saulson, beccaria, 
hughesthorne}. The usual region of concern are frequencies
around $10~ Hz$, where the expected noise can be dangerous 
for the planned advanced g.w. interferometers. In contrast, we are 
concerned with perturbations at much lower frequencies, 
$\sim 10^{-4}~ Hz$, and with their effect on the control 
systems of the already operating initial interferometers. 
Some families 
of seismic perturbations are not accompanied by variations 
of matter density and, hence, they are not accompanied by 
variations of local gravity. This sort of perturbations cannot 
directly affect the plumb lines. However, all seismic 
perturbations can change the distance between the points on 
the Earth surface.  

We will extrapolate the often used spectra of the 
seismic (displacement) noise to the frequencies of 
geophysical interest. For a discussion of seismic noises, 
including those that were measured at the interferometer 
cites, see, for example, \cite{saulson, beccaria, hughesthorne,
AR, angew, 22b, daw}. The displacement noise is normally 
presented in terms of spectral
density $u_f$ written in units of $cm/\sqrt{Hz}$: 
$u_f = A f^{-\gamma}$. The mean-square value of the random
quantity $u$ within a bandwidth from $f_{min}$ to $f_{max}$ is
defined by
\begin{equation}
\label{meansq}
\langle u^2 \rangle = \int_{f_{min}}^{f_{max}} (u_f)^2 df.
\end{equation}    
The spectral index $\gamma$ vary from one spectral interval 
to another, but there exist indications that 
$\gamma \simeq 1$ at frequencies below $0.1~ Hz$, so we will adopt 
$\gamma = 1$. The coefficient $A$ is also uncertain. It 
varies from site to site, and there exist a broad range of
estimates. We will adopt some averaged value 
$A = 10^{-4}~cm /\sqrt{Hz}$ at $f < 0.1~Hz$. So, we shall be 
working with the spectrum   
\begin{equation} 
\label{seism}
u_f = 10^{-4} \left( \frac{10^{-1} Hz}{f}\right) ~\frac{cm}{\sqrt{Hz}}.   
\end{equation}
Surely, at very low frequencies, the concept 
of seismic perturbations overlaps with the concept of the Earth 
free oscillations, so the estimates should be consistent with each 
other and with the gravimeter records. 

To find the differential noise spectral density $\Delta x_{seism}$, 
we have to multiply Eq.(\ref{seism})
with $L/ \lambda = L f/ c_s$. Assuming that $c_s$ does not 
significantly depend on $f$, the frequency dependence cancels out
in the product, 
and the amount of seismic noise is expected to be approximately
equal at all geophysical frequencies. We want to use one simple formula
in the broad interval of frequencies, so we put $c_s = 1~ km/sec$
for this calculation. Then, using $L = 3~ km$, we arrive at 
\begin{equation}
\label{seism2}
\Delta x_{seism} \approx 3 \times 10^{-5}~ \frac{cm}{\sqrt{Hz}}.
\end{equation}
This estimate is a factor of 10 lower than the level of tolerable
noise in Eq.(\ref{noises}). It shows that this source of noise cannot
prevent the observation of the Earth inner core oscillations by the 
adjustment system of laser interferometers.

To check the consistency of our evaluations, it is instructive to 
estimate the upper bound of gravity variations caused by seismic 
perturbations. This evaluation of seismic $\delta g_f$ will also 
be helpful in the next subsection where we are considering atmospheric 
fluctuations. We will have to compare our estimations with
the actually measured (total) values 
of the acceleration noise $\delta g_f$ at the core oscillation 
frequencies around $f_C$ and normal mode $mHz$ 
frequencies (compare, for example, with \cite{richetal, 5aa, 18, tu}):
\begin{equation}
\label{gravvar2}
\delta g_f \approx 0.5~ \left(\frac{10^{-4}~ Hz}{f}\right)
~ \frac{\mu Gal}{\sqrt{Hz}}. 
\end{equation}

A matter density wave, with the amplitude of density variations 
$\delta \rho/ \rho$ and wavelength $\lambda$, produces a local
variation of gravity with the amplitude 
$\delta g/ g(\lambda) \sim \delta \rho/\rho \sim u/\lambda$, where
$g(\lambda)/ g \approx \lambda/R_e$. We take into account the finite
rigidity of Earth, expressed by the Love numbers, and write 
$0.3~ \delta g/g(\lambda) \approx u/\lambda$, that is, 
\begin{equation}
\label{gravvar} 
0.3~ \frac{\delta g}{g} \approx \frac{u}{R_e}.
\end{equation}  
Surely, this formula is consistent with Eq.(\ref{tides3}). 
Using Eq.(\ref{seism}) in this formula, we arrive at the estimate 
for seismic $\delta g_f$, which turns out to be at the level of the 
observed variations, Eq.(\ref{gravvar2}).

The estimate (\ref{gravvar2}) can now be translated into the 
spectral noise contribution to $\Delta \alpha_0$ caused by seismic 
perturbations. It is seen from Eq.(\ref{gen}) that
the $f$-dependence cancels out, so we derive 
\begin{equation}
\label{angvar}
\Delta \alpha_{seism} \approx 10^{-13}~ \frac{rad}{\sqrt{Hz}}. 
\end{equation}
This noise is lower that the tolerable level 
of noise in Eq.(\ref{noises}). Therefore, the alighnment system 
of laser interferometers can also be used, barring some complications
discussed below, for observation of the inner core oscillations.

\medskip
\noindent {\it b) Atmospheric noises}
\medskip

It is somewhat surprising that the movement of light air 
in the Earth atmosphere can be a more serious source of 
noise for the proposed measurements than the 
movement of heavy soil in seismic perturbations. 
We will illustrate this statement with a simple theory of gravity 
variations and land deformations caused by atmospheric 
motions. 

Consider an air density inhomegeneity with the amplitude 
$\delta \rho/\rho$ and a characteristic spatial scale $\lambda$. 
Let this scale be shorter than the characteristic height of the 
Earth atmosphere $H$. In a point of observation at the Earth surface, 
the variation of $g$ is mostly determined by the variation of mass 
$\delta m$ in a volume $\lambda^3$ with the center located at a 
distance $\lambda$ from the observer. More distant volumes give 
smaller contributions to $\delta g$ and can be ignored. 
So, we could write:
\[
\delta g \approx \frac{G~ \delta m}{\lambda^2} \approx 
G~ \delta \rho~ \lambda.
\]

This expression is adequate for $\lambda < H$. However, we are mostly 
interested in the opposite case $\lambda > H$. As will be clear from 
discussion below, it is time-dependent atmospheric variations at these 
longer scales that fall in the frequency range of geophysical interest. 
Therefore, we have to consider variations of mass in a volume 
$H \lambda^2$, rather than $\lambda^3$. The center of
this flattened volume is located at a distance $\lambda$ 
from the observer. The expression for $\delta g$, appropriate for
scales and frequencies of our interest, should be written as
\begin{equation}
\label{vargatm}
\delta g \approx \frac{G~ \delta m}{\lambda^2} \approx 
G ~\delta \rho ~ H. 
\end{equation}

Variations of air density $\delta \rho$ are more difficult to measure 
than variations of pressure, so one usually operates with the 
(spectral) pressure variations $\delta p_f = c_a^{~2}~ \delta \rho_f$, 
where the square of the air sound speed is 
$c_a^{~2} \simeq 10^{9}~ cm^2/ sec^2$.   
Then, equation (\ref{vargatm}) takes the form  
\begin{equation}
\label{vargatm2}
\delta g_f \approx G~ \frac{\delta p_f}{c_a^{~2}}~ H. 
\end{equation}
The question arises which $\delta p_f$ should be used 
in this equation in order to estimate $\delta g_f$.

The observations indicate \cite{c10, c13, 22a, TA, GGM, tu} 
that, during relatively quiet times, the
spectral pressure variations behave, roughly, 
as $\delta p_f \propto f^{-1}$. This frequency dependence is close 
to the Kolmgorov-Obukhov law for atmospheric turbulence \cite{LL, GGM}
$\delta p_f \propto f^{-7/6}$. In the frequency interval 
$\sim (10^{-5} - 10^{-2})~Hz$, observational data
can be approximately fitted by the formula 
\begin{equation}
\label{KO}
\delta p_f = 200 \left(\frac{f}{10^{-4} Hz}\right)^{-1} 
~\frac{Pa}{\sqrt{Hz}}, 
\end{equation}
where the unit of pressure $Pa$ is $Pa = 10 ~g/cm~ sec^2$. The
replacement of the spectral index $(-1)$ by the genuine 
Kolmogorov-Obukhov index $(-7/6)$ leaves the fit equally good. 

A purely turbulent $\delta p_f$ raises certain concerns, as 
it is not obvious that turbulence can directly contribute to 
$\delta g$. Turbulence describes chaotic velocities in a medium and 
variations of the medium's kinetic pressure, but these variations are 
not necessarily accompanied by variations of the medium's mass 
density $\delta \rho$ and gravity $\delta g$. [There exists,
however, an indirect influence of air turbulence on $g$ 
through the land deformations excited by the variable pressure.
We will discuss this issue separately, at the end of this 
subsection.] On the other hand, transportation and mixing of cold 
and hot air, caused by the turbulence, does contribute 
to $\delta \rho$ and, hence, directly contributes to $\delta g$. 
Without analysing the true nature of the observed $\delta p_f$,
we will be using the observed spectrum (\ref{KO}) in
the expression (\ref{vargatm2}). 

Combining Eq.(\ref{KO}) and Eq.(\ref{vargatm2}), and 
using numerical values of the participating quantities,
including the effective height of the atmosphere $H \approx 10~km$, 
we arrive at
\begin{equation}
\label{gravvara}
\delta g_f \approx 0.2~ \left(\frac{10^{-4} Hz}{f}\right) 
~ \frac{\mu Gal}{\sqrt{Hz}}. 
\end{equation}
This evaluated $\delta g_f$ is smaller than, and is consistent with, 
the observed (total) variations $\delta g_f$ approximated by 
Eq.(\ref{gravvar2}).

To evaluate the contribution of the atmospheric $\delta g_f$ to
the noises in distance and angular measurements 
(see Eq.(\ref{gen})), we first need to find out the characteristic
correlation length $\lambda$, as a function of frequency $f$.
To do this, we have used experimental data which were kindly 
provided to us by the Japanese Weather Association. These data
were collected by the Meteorological National Geographical
Institute at the set of meteo-stations near Tsukubo Scientific Center.
The total area of $20 \times 20~km$ is covered by $10$ stations.
The average distance between nearby stations is $10~km$. The pressure
data were recorded at each station with a sample time $1~min$. The
length of the analysed record was $1~month$.  

Using the experimental data, we have built temporal spectra of pressure
$S_p(f)$ measured at each station, and also temporal spectra of difference 
of pressure $S_{\Delta p} (f)$ at each pair of neighbouring stations. 
Because of the spatial correlation of the data, the spectral components
of the difference of pressure $S_{\Delta p} (f)$ prove to be smaller 
than the approximately equal spectral components of pressure $S_p(f)$
measured at each station. 
For each pair of stations, the spatial correlation scale $\lambda(f)$ 
was estimated from the relationship 
\[
\lambda (f) = r \frac{2 S_p(f)}{S_{\Delta p}(f)}.
\]
Then, the found quantities $\lambda (f)$ were avaraged over all pairs
of stations. As a result, we have arrived at the mean value of 
$\lambda (f)$:
\begin{equation}
\label{lamf}
\lambda (f) = 500~ \left( \frac{10^{-4} Hz}{f} \right)~km.
\end{equation}
It also follows from the data that this number depends on overall 
meteorological conditions and can change by a factor $2-3$. 

The evaluations (\ref{lamf}), (\ref{gravvara}) should now be used in 
general formulas (\ref{gen}). We arrive at the expected atmospheric 
noise contribution:
\begin{equation}
\label{atmn}  
{\Delta x}_{atm} \approx 2 \times 10^{-4}~ \frac{cm}{\sqrt{Hz}}, ~~~~~
{\rm and}~~~~ 
{\Delta \alpha}_{atm} \approx 10^{-12}~ \frac{rad}{\sqrt{Hz}}. 
\end{equation}
These estimates are marginally consistent with the level of allowed 
noise defined by Eq.(\ref{noises}).

It is satisfying that the differential displacement noise (\ref{atmn}) 
is consistent with the direct estimate of land deformations, supposedly 
excited by atmospheric turbulence \cite{tu}. The permanently present  
amplitudes of acceleration $a_n$ at $mHz$ frequencies of the Earth's 
eigen-modes are all approximately at the level of $0.4~ nGal$ \cite{tu}. 
Taking $3~ mHz$ as the central frequency, the discrete set of
these observed modal amplitudes $a_n$ can be translated into an
interval of a continuous acceleration spectrum 
\begin{equation}
\label{acc}
a_f \approx 0.4~ \frac{1}{\sqrt{3\times 10^{-3}}}~ \frac{nGal}{\sqrt{Hz}}
\approx 7~ \frac{nGal}{\sqrt{Hz}}.
\end{equation}
The observed accelerations $a_f$ are associated with the 
periodic displacements $u_f$ of the land elements: 
\begin{equation}
\label{displ}
u_f \approx \frac{a_f}{(2 \pi)^2 f^2} \approx 10^{-9}~ \frac{1}{f^2}
~\frac{cm}{\sqrt{Hz}}.
\end{equation}
The corresponding spectral noise amplitude of the distance variation 
between the interferometer's mirrors, separated by $L= 3~km$, will 
amount to
\begin{equation}
\label{tun}
\Delta x_{f} \approx u_f \frac{Lf}{c_s} \approx 3 \times 10^{-9} 
~\frac{1}{f}~ \frac{cm}{\sqrt{Hz}}.
\end{equation}
We extrapolate this formula to lower frequencies, including
the interval around $f_C$. Then, at $f=f_C$, we obtain  
\begin{equation}
\label{tun2}
\Delta x_{f} \approx 5 \times 10^{-5}~ \frac{cm}{\sqrt{Hz}}.
\end{equation}
This estimate is perfectly consistent with the independent evaluation 
that we have derived in Eq.(\ref{atmn}). 

To conclude this subsection, we can say that the expected 
atmospheric noise is marginally acceptable for the proposed 
geophysical measurements with laser-beam detectors of gravitational 
waves. It is likely that some rare events, such as passing  
cyclones and atmospheric fronts, can produce larger disturbances than
the estimates (\ref{atmn}), but these atmospheric events can be 
anticipated and excluded from the data.    

\medskip
\noindent {\it c) Coupling of noises in distance and angle measurements}
\medskip

As was already discussed above, variations of the Earth's 
gravitational field, caused either by external or internal
sources, produce changes in, both, distance between the sites 
and angle between the plumb line directions on the Earth surface. 
It is only on the surface of an idealized perfectly 
rigid body that the change of angle between the plumb lines is 
not accompanied by distance variations. And it is only
in an idealized homogeneous gravitational field, or during 
spherically-symmetric oscillations of a homogeneous sphere, 
that the change of distance between the sites is not accompanied by the 
change of angle between the plumb lines. While describing geophysical 
signals, we have already noticed this important link between 
$\Delta x_0$ and $\Delta \alpha_0$, see Eq.(\ref{coupl}). 
For example, as we know, the Earth inner core oscillations
can lead to distance variations with the amplitude 
$\Delta x_0 = 3 \times 10^{-7}~ cm$, and they will necessarily 
be accompanied by the angle variations with the amplitude
$\Delta \alpha_0 = 2 \times 10^{-15}~ rad.$ However,
even in the absence of any geophysical signal, a similar link
exists between the random (noisy) parts of $\Delta x_0$ and 
$\Delta \alpha_0$.

\begin{figure}
  \begin{center}
\includegraphics{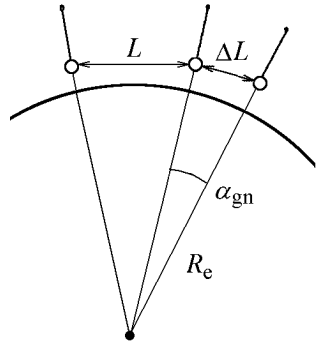}
  \end{center}
 \caption{\label{f7} Coupling of signals and noises in distances and angles}
\end{figure}

Fig.\ref{f7} illustrates the appearance of noise in the
channel of angle measurements, if there exists noise in 
the channel of distance measurements. The fluctuating part
$\Delta L$ of distance between the sites (independently of
the reasons why this fluctuating part exists) translates into the 
fluctuating part $\alpha_{gn}$ of angle between the plumb
lines, $\alpha_{gn} \sim \Delta L/ R_e$. This coupling takes
place simply because of the centrally-symmetric character 
of the Earth's gravitational field. [We ignored this fact
in derivation of Eq.(\ref{9}), but otherwise a term proportional 
to $\Delta x_{0 \omega}/ R_e$ should have been added to the 
right-hand-side of that equation.] Therefore, signals and 
noises in the two channels are linked by essentially one 
and the same relationship: 
\begin{equation}
\label{crel}
\Delta \alpha \approx \frac{\Delta x}{R_e}. 
\end{equation}
This equation gives the minimal and unavoidable level of noise in 
angles, which arises because of the noise in distances. Other reasons 
(for example, related to local inhomogeneities on two sites) can 
only make this relationship more complicated. However, even the simplified 
link represented by Eq.(\ref{crel}) leads to an important conclusion: 
there is no special advantage in measuring angles through the
alignment control system, rather than measuring distances
through the adjustment control system; the signal-to-noise ratios 
are practically equal in these channels. On the other hand, the use of
both channels, instead of only one of them, would certainly
increase the reliability of geophysical detection. In particular,
the combined information will help identify and remove the distance
variations caused by
quasi-periodic thermal deformations of the Earth surface. Moreover, 
if both the distance and angles between mirrors are
accurately measured (including their noisy contributions), 
then the use of Eq.(\ref{crel}) for noises (or, preferably, the 
use of a more precise version of this equation) would allow  
one to subtract from the measured angle that part of the angle 
which arises purely due to the coupling. This procedure would help
identify that part of the measured angle which is caused specifically 
by the variation of plumb lines.

It turns out, however, that the angle measurements by the 
actually existing alignment systems have their own difficulties
and ambiguities. We will fully address them in Sec.5. 

\medskip
\noindent {\it d) Instrumental noises}
\medskip

It appears that instrumental noises will not be a major problem for
the proposed geophysical studies. This is not surprising, as the 
gravitational-wave detectors implement the best currently available
technology aimed at measuring much weaker gravitational signals.
Obviously, geophysical signals have different origin and 
are concentrated in a substantially different frequency band, but
in both cases we are dealing with manifestations of the same
force - gravitation. 

The tolerable noise of Eq.(\ref{noises}) implies that the
accuracy of the adjustment system is better
than $3 \times 10^{-6}~ cm$ at frequencies around $10^{-4}~ Hz$.   
Adjustment systems of the presently operating interferometers
satisfy this requirement, but, in order to ensure a successful 
performance of the interferometer as a g.w. detector, they may 
not need the low-frequency precision higher than $10^{-7}~ cm$.  
Certainly, the advanced interferometers will require an enhanced
precision of the adjustment systems in the `control band' of 
frequencies. This is necessary in order to 
assure that only very small forces 
are applied to the mirrors themselves \cite{sho}. In any case, 
from the instrumental point of view, the accuracy of the 
adjustment system can be made much better than presently achieved. 
Even smaller variations, at the level of $10^{-13} cm$, can be 
detected by the interferometer adjustment system, if the dominant
noise is reduced to the photon shot noise \cite{MT1,MT2}. 
The photon noise limited spectral sensitivity is basically determined 
by the relationship
\begin{equation}
\label{Ed3}
\Delta x_{instr} \simeq
\frac{\lambda}{F} \sqrt{\frac{\hbar\omega}{\eta P}},
\end{equation}
where $\lambda$ is the wavelength of the laser light and
$\omega= 2\pi c/\lambda$ is its frequency, $F$ is the 
cavity finesse, $P$ is the optical power available at the photodector
and $\eta$ is its efficiency. Taking reasonable parameters
$\lambda = 10^{-4}~ cm$, $F = 100$, $P = 100~ mW$, $\eta =0.9$
we arrive at  
\begin{equation}
\label{ed3a}
\Delta x_{instr} \approx 1.5 \times 10^{-15} ~~\frac{cm}{\sqrt{Hz}}. 
\end{equation}
This number can be improved by the increase of $P$. Therefore, 
we do not expect instrumental noises in
the adjustment system to be a limiting factor for the measurement
of geophysical signals described by Eqs. (\ref{oscil}), (\ref{core}).
Obviously, we assume that the extra noises (above purely photon noise)
are not excessively large.

The purpose of the alignment system is to monitor and remove
possible misalignments between the laser beam and the 
axis of optical resonator formed by the corner mirror and the end
mirror. One reason for this misalignment to arise is precisely 
the deviation of plumb lines caused by a geophysical signal and, 
hence, the change of angle between mirrors. In principle, the existing
alignment systems are very sensitive to a
possible change of angle between mirrors. If the dominant noise
is the photon shot noise, the angular spectral precision of the
alignment system can reach \cite{8, 9, 1a, 9a}: 
\begin{equation}
\label{E3}
\Delta \alpha_{instr} \simeq
\frac{\lambda}{w_0} \sqrt{\frac{\hbar\omega}{\eta P}},
\end{equation}
where $w_0$ is the diameter of the laser beam `waist'.
Taking the same parameters as before and $w_0= 0.1~ cm$, 
we arrive at  
\begin{equation}
\label{e3a}
\Delta\alpha_{instr} \approx 1.5 \times 10^{-12} ~~\frac{rad}{\sqrt{Hz}}. 
\end{equation}
This accuracy is at the level of the tolerable noise of Eq.(\ref{noises}),
but it can be improved by the increase of $P$. We cautiously conclude that 
the instrumental noise of alignment systems will not swamp the expected 
geophysical signal. 

A more fundamental problem with the existing alignment systems 
lies, however, in a different place. The error signal of the system 
cannot tell us whether the misalignment arose because of the change of 
angle between the mirrors or because of the change of lateral 
position of the end spherical mirror. The first cause is
what we are interested in, while the second cause could be simply 
an unrelated noise. Without solving this problem, one would not be able
to use angular measurements as an additional source of geophysical 
information. This is why we address this problem in next sections.

\section{Measurement of mirrors' tilts}

Here, we will consider in more detail the response of a g.w.
interferometer to the deformations of Earth's surface and 
variations of local gravity. To be specific, we will do this
on the example of the VIRGO observatory \cite{virgoweb}.  

Each arm of the VIRGO interferometer consists of an asymmetric
Fabry-Perot optical resonator, which is formed by a flat front 
mirror and a spherical end mirror. The length of each arm is 
$L=3~km$. The laser beam enters the resonator through the partially 
transparent front mirror. Each mirror is hanging on a multi-stage
support, which behaves as a single wire pendulum at frequencies
below $1~Hz$. The source of light itself is located very close 
(at a distance less than $20~m$) to the front mirror, so one can
neglect the gradients of local gravity and deformations of land at
such short distances. Therefore, we shall accept, for simplicity, 
that in the chosen coordinate system the front flat mirror does not 
move, the laser beam is always orthogonal to the front mirror, and the
beam enters the resonator from one and the same point on the flat mirror. 
The problem is still sufficiently general, as the main trouble arises 
when one considers the relative position and orientation of the remote 
spherical mirror with respect to the front mirror and laser beam.     

\begin{figure}
  \begin{center}
\includegraphics{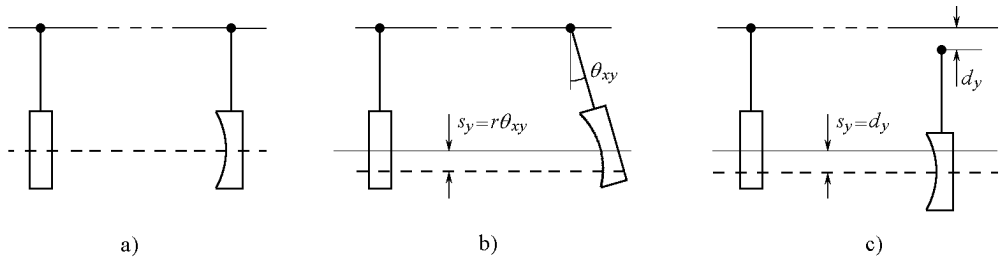}
  \end{center}
 \caption{\label{mirrors} Misalignment of the optical axis in 
the Fabry-Perot interferometer: a)~aligned configuration, 
b) tilt of the spherical mirror, c) vertical shift of the 
spherical mirror}
\end{figure}

The optical axis of the resonator is orthogonal to the flat mirror
and goes through the center of curvature of the spherical mirror.
The curvature radius $r$ of the spherical mirror is approximately
equal to the distance between the mirrors, so we take $r= 3~km$.
In the tuned working condition, the laser beam is aligned with 
the optical axis. In this configuration, all light participates 
in interference, and the sensitivity is maximal. A misalignment 
between the beam and optical axis is sensed by the changing spatial 
distribution of the electromagnetic field in the Fabry-Perot 
cavity \cite{8, ward}. Since we have assumed that the
beam is always orthogonal to the flat mirror, the misalignment
can only arise if the spherical mirror inclines with respect to
the beam, or changes its position in the plane orthogonal to 
the beam. These possible 
motions of the end mirror, in the $(x, y)$-plane containing the 
beam, are shown in Fig.\ref{mirrors}.    

A technical advantage of this alignment control system (which turns 
out to be a big disadvantage for geophysical applications) is in 
that the system does not need to know whether the misalignment
arose because of the tilt of the end mirror or because of its lateral
shift. They both lead to similar displacements of the optical axis, 
which will be corrected by the control system without ever
distinguishing the true origin of the displacement. In other 
words, there exists a degeneracy between tilts (configuration b) 
in Fig.\ref{mirrors}) and shifts
(configuration c) in Fig.\ref{mirrors}). They both can
result in one and the same parallel displacement of the optical 
axis (shown by a dashed line in Fig.\ref{mirrors}). The 
error signal automatically arises when a misalignment develops, 
but it will be the same signal for both configurations. In both cases, 
the feedback mechanical system compensates the misalignment by 
tilting the spherical mirror and returning the optical axis back 
to its original position aligned with the laser beam.
 
Let the $x$-axis go along the beam and the $y$-axis be a vertical
direction. Suppose that the plumb line tilts by an angle 
$\theta_{xy}$  (see Fig.\ref{mirrors}). Then the curvature center
of the hanging spherical mirror moves a distance 
$r \theta_{xy}$. This means that the optical axis will take a new
position displaced from the original one by $s_y = r \theta_{xy}$. 
Now let the suspension point of the spherical mirror shift by 
a distance $d_y$ (see Fig.\ref{mirrors}). Then the optical axis does 
also shift by this distance, $s_y=d_y$. The parallel displacements 
of the optical axis will be numerically the same, 
if $d_y =r \theta_{xy}$.

The expected geophysical signal in angle variations is
$\Delta \alpha = 2 \times 10^{-15}~rad$, see Eq.(\ref{core}). 
The displacement of the optical axis caused by this
useful signal will be $s_y = 6 \times 10^{-10}~ cm$. This is a
measurable signal. However, such a displacement of the axis 
can also be produced by a lateral shift of the suspension point 
$d_y = 6 \times 10^{-10}~ cm$. These lateral shifts of the suspension
point are likely to happen  - they are well below the 
displacement noise of Eq.(\ref{noises}). All spatial components of
the displacement noise have approximately equal amplitudes,
so one can be sure that there is also a vertical component 
of this level of magnitude. 
In other words, a displacement noise, which is tolerable for 
longitudinal distance measurements, turns out to be intolerable
for angular measurements by the existing alignment systems.
Although the control system will generate an error signal and
will correct the misalignment, it is impossible to distinguish
whether it was produced by a useful geophysical effect or by the 
ever-present displacement noise. 

Since the channel of angular measurements is important
for geophysical applications, we have to find a way of breaking the
degeneracy, specific for spherical mirrors, between tilts and 
shifts. This will require modest
amendments to the existing optical configurations. Some possible
ideas are described in the next section. It is also important to   
note that the advanced interferometers may not need these 
amendments at all. There exist strong arguments in favor of using
non-spherical mirrors \cite{bonthorne} in which case the problem of 
degeneracy will be automatically alleviated.

\section{Modifications of the optical scheme}

It was shown above that the useful geophysical information on 
$\Delta x_0$ is routinely stored in circuits of the adjustment 
system. The extracton of this part of geophysical information does
not require any hardware modifications, and it can be found at the 
level of data analysis. In contrast, to receive appropriate information 
on $\Delta \alpha_0$ one will first need to make certain modifications 
of g.w. detectors with spherical mirrors, specially for the purpose of 
geophysical applications. In principle, the necessary 
angular information could be obtained by methods unrelated 
to g.w. observations, but the modifications proposed here make
use of the existing exceptional infrastructure of g.w. observatories, 
such as long arms incorporating high-quality vacuum tubes, 
some common for astrophysical and geophysical applications elements of 
shielding and control, sensitive detection techniques, etc. Surely, 
we keep in mind that any geophysical modifications should not jeopardize 
the functioning of interferometers as astrophysical instruments.  

\subsection{Auxiliary interferometer for distance measurements}

One way of receiving information on $\Delta \alpha_0$ is based on
Eq.(\ref{10}). The adjustment system responds to the low-frequency part of 
the distance variation $\Delta d_{\omega}$ between mirrors. 
As we know, the major contribution to $\Delta d$ is provided 
by $\Delta x_0$, and this is why we intend 
to extract information on $\Delta x_0$ from the data of the 
adjustment system. However, the accuracy of this measurement can be so 
high that it can reach the level of a much smaller contribution to 
$\Delta d$: $l \Delta \alpha_0$, second term in Eq.(\ref{10}). 
Taking $l=10^{2}~ cm$ for
the suspension length and $\Delta \alpha_0 = 2 \times 10^{-15}~ rad$,  
Eq.(\ref{core}), for the expected useful signal, we get the 
estimate: $l \Delta \alpha_0 \approx 2 \times 10^{-13}~ cm$.  
If one could measure the difference between $\Delta d$ and $\Delta x_0$ 
with this sort of precision, and since the suspension length $l$
is known, the result would have provided the necessary 
information on $\Delta \alpha_0$. In principle, a direct measurement 
of this difference is possible with the help of an additional 
interferometer system, which we will now describe.
This proposal is similar to the one first suggested by R. Drever
in the context of an interferometer on magnetically suspended 
mirrors \cite{DR, DA}.

\begin{figure}
  \begin{center}
\includegraphics{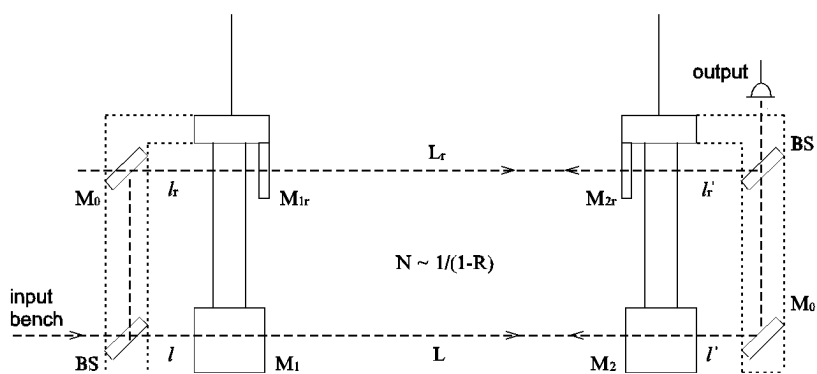}
  \end{center}
 \caption{\label{f4}  Auxiliary  interferometer. M$_1$ and M$_2$ are 
mirrors of g.w. interferometer, M$_{1r}$ and M$_{2r}$ are additional 
mirrors forming the extra resonator, BS and M$_0$ are beamsplitters 
and mirrors changing the beam direction.}
\end{figure}

The auxiliary interferometer is shown in Fig.\ref{f4}. One arm is
formed by mirrors of the g.w. interferometer itself. Another arm
is formed by mirrors attached to the suspension points, or, more
practical, to the last stage of the anti-seismic filters. Such 
configuration with two parallel arms is known as Mach-Zender 
interferometer. Each arm of the Mach-Zender interferometer can 
include a Fabry-Perot resonator.
It is supposed that a small fraction of light is deviated from the 
main beam to the additional interferometer. As the frequencies of 
geophysical interest are relatively low, $\sim 10^{-4} Hz$, the 
relaxation time of the Mach-Zender interferometer should be sufficiently 
long. If the reflectivity of mirrors $M_{1r}$ and $M_{2r}$
is sufficiently high, a small portion of light 
can resonate in the cavity for long time, increasing the sensitivity of
observation. The added interferometer is capable of measuring
the difference of its arm-lengths with high precision.  
It is interesting to note that this configuration has been actually 
realised \cite{tsu}, even though the modification is motivated by the
struggle with noises rather than by geophysical applications. 

The proposed scheme is differential, and therefore it has certain 
advantages:

- seismic and other common for both arms noises cancel out due to 
  differential character of measurements;

- sudden changes of laser beam position and direction (beam `walks' 
  and `jitter') are not dangerous, as they are common
  for both arms and cancel out; 

- the scheme is universal and can be used at any g.w. interferometer 
  regardless of the actual construction of suspension.

Two other possible modifications, discussed next, are aimed at 
direct breaking of degeneracy between shifts and tilts in 
the alignment system. These modificatons do not imply a deviation of
light from the main beam of interferometers and, therefore, 
they are even less likely to compromise the g.w. performance of the 
instruments.

\subsection{\label{sfl} Flat mirror behind spherical end mirror.}

\begin{figure}
  \begin{center}
\includegraphics{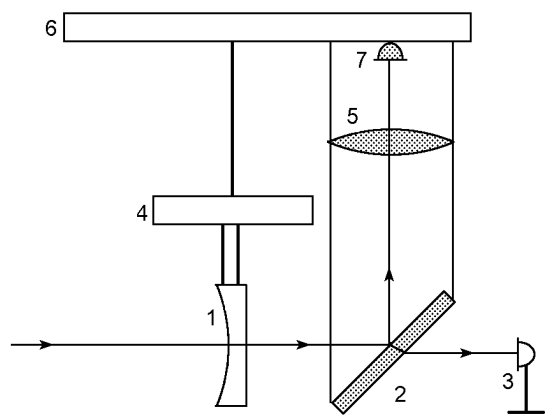}
  \end{center}
 \caption{\label{f5} A flat mirror (beamsplitter) (2) behind the spherical 
end mirror (1). A photodetector of the main alignment system is 
denoted (3), the focusing lens - (5), the last stage of anti-seismic 
filter - (6), the additional photodetector - (7).}
\end{figure}

The desired goal of telling the difference between shifts and tilts 
of spherical end mirror can be achieved by installation of a 
flat beamsplitter behind the end mirror. The 
beamsplitter can be attached to the last stage of the anti-seismic
filter. A scheme of this proposal is shown in Fig.\ref{f5}. It is 
supposed that the beamsplitter (2) intercepts part of light traveling 
toward the photodetector (3), normally used as an element of the
alignment system. The intercepted light is directed through the focusing
lens (5) toward the additional photodetector (7). This photodetector should
be sensitive to the position of a spot of light focused on the photodetector.
For example, the photodetector (7) could consist of two or four pieces,
so that a slight movement of the focal spot would produce a differential
signal. Since the spherical mirror (1) and the beamsplitter (2) are
hanging very close to each other, they participate together in possible 
shifts of the suspension point and in possible tilts caused by varying 
local gravity.   

Imagine that the mirror (1) and beamsplitter (2) are shifted together in 
vertical direction (configuration c) in Fig.\ref{mirrors}). This would
lead to no change at the photodetector (7). In contrast, if the mirror (1) 
and beamsplitter (2) are tilted together (configuration b) in 
Fig.\ref{mirrors}), the focal spot will change its position on the     
photodetector (7) leading to a measurable signal. In this way, the
previously indistinguishable configurations b) and c) become 
distinguishable. If the evaluation (\ref{e3a}) of the angular precision 
remains valid for the proposed scheme, one will be able to extract 
geophysical information on $\Delta \alpha_0$ from this measurement.

\subsection{\label{sla} Additional laser behind the end mirror.}

\begin{figure}
  \begin{center}
\includegraphics{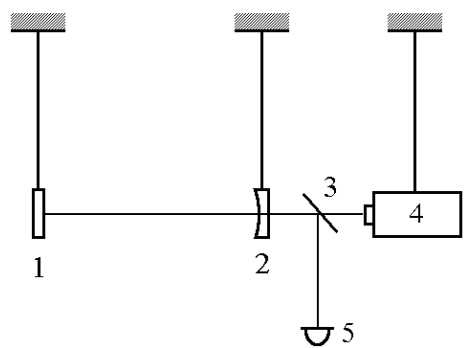}
  \end{center}
  \caption{\label{f6}  Additional laser behind the end mirror: (1) and
(2) - the front and the end mirrors of the main interferometer; 
(3) - beamsplitter; (4) - additional laser; (5) - additional photodetector.}
\end{figure}

The most natural solution to the formulated problem is, perhaps, the 
installation of an additional source of light (for some extra details on this 
proposal, see \cite{BBG}). Imagine that the additional laser is suspended 
(possibly, from the same anti-seismic filter) behind the spherical end 
mirror and shines inside the main cavity, see Fig.\ref{f6}. In order to 
avoid any mixing of the new light with the already present light in the 
cavity, the frequency of the auxiliary laser should differ from the frequency
of the main laser. However, inside the cavity, the added light satisfies all 
the usual requirements on mode matching, alignment of the light beam with the 
optical axis, etc. A possible misalignment of the new beam with the optical 
axis of the Fabry-Perot resonator will be sensed by the same alignment system 
and in exactly the same manner as it takes place for the main beam. The only 
difference is that the detected misalignment of a new beam will be recorded 
for geophysical studies rather than used for correcting the inclination of 
the end mirror. One will also need a new photodetector for the output light.  

The auxiliary laser and the end mirror are hanging very close to each other, so
they react in the same way to shifts of the suspension point or tilts of the 
plumb line. If the laser itself and the end mirror are both subject to a shift 
of the suspension point (configuration c) in Fig.\ref{mirrors}) the alignment 
of the additional beam with the optical axis of the Fabry-Perot resonator 
remains intact. The system does not generate an error signal. In contrast, 
a tilt of the plumb line (configuration b) in Fig.\ref{mirrors}) will 
cause a tilt of direction 
of the incoming beam of the auxiliary laser. This tilt will be detected by the 
alignment system, thus distinguishing the configurations b) and c). The arising 
error signal is what we are after. Assuming that the angular sensitivity is 
still determined by Eq.(\ref{e3a}), one will get access to
the information on $\Delta \alpha_0$. In fact, the installation of an extra 
laser with properties similar to the ones described above was originally 
discussed in a different context, with the aim of facilitating the initial 
tuning of the main interferometer.

\section{Conclusions}

The need for successful functioning of laser-beam detectors of 
gravitational waves inevitably makes them valuable 
geophysical instruments. The most immediate application of g.w. 
detectors is accurate measurement of low-frequency distance
variations between the central mirror and the end 
mirrors. Essentially, the adjustment control system makes the 
interferometer a low-frequency 3-point meter of deformations. The 
principal advantage of the existing g.w. detectors in comparison 
with traditional geophysical techniques is exceptionally
long arms of interferometers (a few kilometers in comparison with 
typical hundred-meter long geophysical interferometers) and high 
sensitivity of measurements. This advantage leads to a much better
strain sensitivity. The extraction of deformational part of the 
geophysical signal does not imply any hardware modifications, and 
it can be found in records of the adjustment system right at the 
level of data analysis.

The alignment system of g.w. laser interferometers on suspended
mirrors can potentially provide even more interesting geophysical 
information - the relative variation of local gravity vectors 
(plumb lines). This sort of information cannot be obtained from 
geophysical interferometers which normally use mirrors attached 
to the ground. It appears, however, that the extraction of this 
information from the existing g.w. laser interferometers will require
certain modest modifications of their optical scheme. Without 
compromising the astrophysical program of the instruments, these 
modifications seem to be possible, and we suggested several ways 
of doing this. Their realization will allow one to use angular 
measurements in addition to distance measurements. In future
advanced g.w. interferometers based on non-spherical mirrors these
modifications may be avoided.

The evaluation of interesting geophysical signals and 
comparison with environmental and instrumental noises shows
that the g.w. instruments can reach the most ambitious geophysical
goals, such for example as detection of oscillations of the inner 
core of Earth. 

It seems that further progress in this area can be achieved, 
if a closer collaboration is established between geophysicists and 
representatives of g.w. community. 

\section*{Acknowledgments} 

At various stages of this research we have benefited from discussions
with P. Bender, R.~Drever, J. Hough, A. Kopaev, K. Strain, K. Tsubono, 
H. Ward and, especially, A.~Giazotto. This work was partially 
supported by a grant from the Royal Society (UK) for international 
collaboration, RCPX331.  

\vspace{1cm}

{\it Note added:} In the very last days of preparation of this paper
for publication we have tragically lost our young talanted colleague
and friend - Andrej Serdobolski.

\end{document}